\documentstyle[12pt,fleqn]{article}
\def\soc{{\rm C}_{60}}
\def\rug{{\rm C}_{70}}

\def\beeq{\begin{equation}}
\def\eneq{\end{equation}}
\def\beeqa{\begin{eqnarray}}
\def\eneqa{\end{eqnarray}}

\addtocounter{section}{-1}
\begin{document}

\begin{center}

\vspace{2in}

{\large {\bf{Metal-Insulator Transition in C$_{\bf 60}$-Polymers
} } }

\vspace{1cm}

{\rm Kikuo Harigaya\footnote[1]{E-mail address:
harigaya@etl.go.jp; URL: http://www.etl.go.jp/People/harigaya/.}
}\\

\vspace{1cm}

{\sl Fundamental Physics Section,\\
Electrotechnical Laboratory,\\
Umezono 1-1-4, Tsukuba, Ibaraki 305, Japan}

\vspace{1cm}

(Received~~~~~~~~~~~~~~~~~~~~~~~~~~~~~~~~~~~)
\end{center}

\Roman{table}

\vspace{1cm}

\noindent
{\bf Abstract}\\
Variations in the band structures of $\soc$-polymers are
studied, when $\pi$-con\-ju\-ga\-tion conditions are changed.
We look at band structures in order to discuss a metal-insulator
transition, using a semi-empirical model with
the Su-Schrieffer-Heeger type electron-phonon interactions.
We find that electronic structures change among direct-gap insulators and
the metal, depending on the degree of $\pi$-conjugations.
High pressure experiments could observe such pressure-induced
metal-insulator transitions.

\mbox{}


\pagebreak

Recently, it has been found that the linear $\soc$-polymer
is realized in alkali-metal doped $\soc$ crystals:
$A_1\soc$ ($A=$K, Rb) [1-4], and much attention has been
focused on their solid state properties.  One
electron per one $\soc$ is doped in the polymer chain.
It seems that Fermi surfaces exist in high temperatures,
but the system shows antiferromagnetic correlations in
low temperatures [1].  The structure of the $\soc$-polymer
is displayed in Fig. 1.  The $\soc$ molecules are arrayed
in a linear chain.  The bonds between $\soc$ are formed
by the [2+2] cycloaddition mechanism.

Several calculations of the electronic structures have
been performed.  For example, a tight-binding calculation
of a linear chain [5] has been reported, and the relation
to the antiferromagnetic ground state has been discussed.
A semi-empirical tight-binding model [6] analogous to the
Su-Schrieffer-Heeger (SSH) model [7] of conjugated polymers
has been proposed, and the possibility of the charge density
wave state has been pointed out.  The band calculation
by the first principle method has also been done [8].
The electronic structures can become three dimensional
when distances between $\soc$-polymer chains are short,
while they remain one dimensional when the distances are longer.

The above works have been focused upon electronic structures
of the $\soc$-polymer doped with one electron per one $\soc$.
The electronic structures may depend sensitively upon the $\pi$-conjugation
conditions {\sl even in the neutral polymer}, because the several bonds
connecting neighboring molecules are largely distorted and
the mixing between $\sigma$- and $\pi$-orbitals will change
only by slight change of the bond structures [2].   We would
like to study effects of the change of the $\pi$-conjugation
conditions by introducing a phenomenological parameter
in a tight-binding model.  The model is an extension of the
SSH-type model which has been applied to $\soc$ [9,10] and
$\rug$ [10,11] molecules. We look at band structures, in order
to discuss a metal-insulator transition in neutral systems.

In the previous works [9-11], we have proposed the extended SSH model
to $\soc$ and $\rug$.  In $\soc$, all the carbon atoms are
equivalent, so it is a good approximation to neglect the
mixings between $\pi$- and $\sigma$-orbitals.  The presence of
the dimerization and the energy level structures of the neutral
$\soc$ molecule can be quantitatively described by the calculations
within the adiabatic approximation.  In $\rug$, the molecular structure
becomes longer, meaning that the degree of the mixings between $\pi$-
and $\sigma$-characters are different depending on carbon sites.
In this respect, the extended SSH model does not take account of
the difference of the mixings.  However, it has been found [10,11] that
qualitative characters of the electronic level structures are
reasonably calculated when the extended SSH model is applied to
the $\rug$.  This is a valid approach because the energy positions
of the $\sigma$-orbitals are deep enough to neglect them
in the first approximation.

In this letter, we assume the same idea that the lattice
structures and the related molecular orbitals of
each $\soc$ molecule in the $\soc$-polymer can be described
by the SSH-type model with the $\pi$-orbitals only.
However, the mixings between the $\pi$- and $\sigma$-orbitals
near the four bonds, $\langle i,j \rangle$ ($i,j=1 - 4$),
shown in Fig. 1 are largely different from those of regions
far from the four bonds.  We shall shed light on this special
character of bondings between neighboring $\soc$.
Electronic structures would be largely affected by changes
of $\pi$-conjugation conditions around the four bonds.
We shall introduce a semi-empirical parameter $a$ as shown
in the following hamiltonian:
\beeqa
H_{\rm pol} &=&  a \sum_{l,\sigma}
\sum_{\langle i,j \rangle = \langle 1,3 \rangle,
\langle 2,4 \rangle} (- t + \alpha y_{l,\langle i,j \rangle} )
( c_{l,i,\sigma}^\dagger c_{l+1,j,\sigma} + {\rm h.c.} ) \\ \nonumber
&+&  (1-a) \sum_{l,\sigma}
\sum_{\langle i,j \rangle = \langle 1,2 \rangle,
\langle 3,4 \rangle} (- t + \alpha y_{l,\langle i,j \rangle} )
( c_{l,i,\sigma}^\dagger c_{l,j,\sigma} + {\rm h.c.} ) \\ \nonumber
&+& \sum_{l,\sigma} \sum_{\langle i,j \rangle = {\rm others}}
(- t + \alpha y_{l,\langle i,j \rangle} )
( c_{l,i,\sigma}^\dagger c_{l,j,\sigma} + {\rm h.c.} ) \\ \nonumber
&+& \frac{K}{2} \sum_i \sum_{\langle i,j \rangle} y_{l,\langle i,j \rangle}^2,
\eneqa
where $t$ is the hopping integral of the system without the
dimerization in the isolated $\soc$ molecule; $\alpha$ is the
electron-phonon coupling constant which changes the hopping
integral linearly with respect to the bond variable
$y_{l,\langle i,j \rangle}$, where $l$ means the $l$th
molecule and $\langle i,j \rangle$ indicates the pair of
the neighboring $i$ and $j$th atoms; the atoms with $i=1 - 4$
are shown by numbers in Fig. 1 and the other $i$ within
$5 \leq i \leq 60$ labels the remaining atoms in the same molecule;
$c_{l,i,\sigma}$ is an annihilation operator of the $\pi$-electron
at the $i$th site of the $l$th molecule with spin $\sigma$;
the sum is taken over the pairs of neighboring atoms;
and the last term with the spring constant $K$
is the harmonic energy of the classical spring simulating the
$\sigma$-bond effects.

As stated before, the parameter $a$ controls the strength of
$\pi$-conjugations in the chain direction.  When $a=1$,
the $\sigma$-bonding between atoms 1 and 2 (and also 3 and 4)
is completely broken and the orbitals would become like $\pi$-orbitals.
The bond between the atoms 1 and 3 (and that between the atoms
2 and 4) becomes $\sigma$-like.   As $a$ becomes smaller,
the $\pi$-conjugation between the neighboring molecule decreases
and the $\soc$ molecule becomes mutually independent.  In other
words, the interactions between molecules become smaller.
When $a=0$, the $\soc$ molecules are completely isolated
each other.  The band structures of the $\soc$-polymer will change
largely depending on the $\pi$-conjugation conditions.
This problem is the central issue of this letter.

In the literature [6], the possibility of charge density wave states
has been taken into account by regarding two molecule pair as a unit cell.
In contrast, our interests are focused on the one-dimensional
band structure in the spatially homogeneous system, so we do
not consider the doubled unit cell.  The present unit cell
consists of one $\soc$ molecule.   Using the lattice periodicity,
we skip the index $l$ of the bond variable $y_{l,\langle i,j \rangle}$.
In other words, all the molecules in the polymer are assumed
to have the same lattice structure.  The bond variables are
determined by using the adiabatic approximation in the real space.
The same numerical iteration method as in [10] is used here.
We will change the parameter, $a$, within $0 \leq a \leq 1.0$.
The other parameters, $t=2.1$eV, $\alpha = 6.0$eV/\AA, and
$K = 52.5$eV/\AA$^2$, give the energy gap 1.904eV and the
difference between the short and bond lengths 0.04557\AA\\
for one $\soc$ molecule.  We shall use this parameter set here.
The labels A-I in Fig. 1 indicate carbon atoms
which are not equivalent with respect to the symmetry.  The model
is solved by keeping this symmetry of the polymer chain.

Now, we discuss band structures of $\pi$-electrons.
Figures 2(a), (b), and (c) display the band structures for
the $\pi$-conjugation parameters, $a=0.5$, 0.8, and 1.0,
respectively.  In each figure, the unit cell is taken as
unity, so the first Brilloune zone extends from $-\pi$ to $\pi$.
Due to the inversion symmetry, only the wavenumber region,
$0 \leq k \leq \pi$, is shown in the figures.  In Fig. 2(a),
the highest fully occupied band is named as ``HOMO", and the lowest
empty band as ``LUMO".  There is an energy gap about 0.8 eV at the
zone center.  The system is a direct gap insulator.
As increasing the parameter $a$, the overlap of the HOMO and LUMO
appears.  This is shown for $a=0.8$ in Fig. 2(b).
There are Fermi surfaces, so the system changes into a metal.
If $a$ increases further, the positions of the HOMO and LUMO of
the smaller $a$ case
are reversed as shown for $a = 1.0$ in Fig. 2(c).  The system becomes
a direct gap insulator again.   The energy gap is at $k=\pi$.

The above variations of the energy gap are summarized in Fig. 3.
The white (black) squares indicate that the system is
a direct gap insulator where there is a energy gap at
$k=0$ ($\pi$).  The crosses indicate the metallic cases.
The energy gap decreases almost linearly for smaller $a$.
The system changes into a metal as $a$ increases, and
finally an energy gap appears again.   As has been discussed
in [2], the
$\pi$-conjugations between the bonds, $\langle 1,2 \rangle$ and
$\langle 3,4 \rangle$, might be weak.  So, we can assume that
the larger parameter $a$ is reasonable for the real $\soc$-polymer.
There would be a good possibility that the realistic $a$ is
in the region where we can expect metallic and insulating behaviors.
Therefore, it would be interesting to do experiments which give
a high pressure to neutral systems in order to change $\pi$-conjugation
conditions.

In summary, we have studied the variations of the band structures of
the $\soc$-polymer.  We have changed $\pi$-con\-ju\-ga\-tion
conditions by a phenomenological parameter.  A semi-empirical model
with SSH-type electron-phonon interactions has been proposed.
Band structures have been shown extensively, in order to
discuss a metal-insulator transition.
We have found that electronic structures change among direct-gap
insulators and the metal, depending on the degree of $\pi$-conjugations.
The high pressure experiments may be able to change $\pi$-conjugation
conditions in the chain direction, and the electronic structure
changes could be observed.

\pagebreak
\begin{flushleft}
{\bf References}
\end{flushleft}

\noindent
$[1]$ Chauvet O, Oszl\`{a}nyi G, Forr\'{o} L, Stephens P W,
Tegze M, Faigel G and J\`{a}nossy A 1994
{\sl Phys. Rev. Lett.} {\bf 72} 2721\\
$[2]$ Stephens P W, Bortel G, Faigel G, Tegze M,
J\`{a}nossy A, Pekker S, Oszlanyi G, and Forr\'{o} L
1994 {\sl Nature} {\bf 370} 636\\
$[3]$ Pekker S, Forr\'{o} L, Mihaly L and J\`{a}nossy A 1994
{\sl Solid State Commun.} {\bf 90} 349\\
$[4]$ Pekker S, J\`{a}nossy A, Mihaly L, Chauvet O,
Carrard M and Forr\'{o} L 1994 {\sl Science} {\bf 265} 1077\\
$[5]$ Tanaka K, Matsuura Y, Oshima Y, Yamabe T,
Asai Y and Tokumoto M 1995 {\sl Solid State Commun.}
{\bf 93} 163\\
$[6]$ Surj\'{a}n P R and N\'{e}meth K 1994 {\sl Solid State Commun.}
{\bf 92} 407\\
$[7]$ Su W P, Schrieffer J R and Heeger A J 1980
{\sl Phys. Rev.} B {\bf 22} 2099\\
$[8]$ Erwin S C, Krishna G V and Mele E J 1995 {\sl Phys. Rev.}
B {\bf 51} 7345\\
$[9]$ Harigaya K 1991 {\sl J. Phys. Soc. Jpn.} {\bf 60} 4001\\
$[10]$ Harigaya K 1992 {\sl Phys. Rev.} B {\bf 45} 13676\\
$[11]$ Harigaya K 1992 {\sl Chem. Phys. Lett.} {\bf 189} 79\\
$[12]$ Iwasa Y, Arima T, Fleming R M, Siegrist T, Zhou O,
Haddon R C, Rothberg L J, Lyons K B, Carter Jr H L, Hebard A F,
Tycko R, Dabbagh G, Krajewski J J, Thomas G A and Yagi T
1994 {\sl Science} {\bf 264} 1570\\

\pagebreak

\begin{flushleft}
{\bf FIGURE CAPTIONS}
\end{flushleft}

\mbox{}

\noindent
Fig. 1.  The crystal structure of the $\soc$ polymer.
The labels, A-I, indicate carbon atoms which
are not equivalent with respect to the symmetry.  The $\pi$-conjugations
along four bonds, which connect carbon atoms with labels, 1-4,
are controled by the phenomenological parameter $a$ in Eq. (1).

\mbox{}

\noindent
Fig. 2.  Band structures of the $\soc$-polymer of the cases
(a) $a = 0.5$, (b) 0.8, and (c) 1.0, respectively.
In (a) and (c), the highest fully occupied band is
named as ``HOMO", and the lowest empty band as ``LUMO".
The lattice constant of the unit cell is taken as unity.

\mbox{}

\noindent
Fig. 3.  The variations of the energy gap plotted against $a$.
The white (black) squares indicate that the system is
a direct gap insulator where there is a energy gap at
$k=0$ ($\pi$).  The crosses are for metallic cases.

\end{document}